\documentstyle[latexsym,amssymb,epsfig,aps,floats,eqsecnum,
preprint,amsfonts]{revtex}
\def\laq{\raise 0.4ex\hbox{$<$}\kern -0.8em\lower 0.62 ex\hbox{$\sim$}}
\def\gaq{\raise 0.4ex\hbox{$>$}\kern -0.7em\lower 0.62 ex\hbox{$\sim$}}

\begin{document}
\draft
\bibliographystyle{unsrt}

\title{Hedgehogs in higher dimensional gravity 
with curvature self-interactions}

\author{Massimo Giovannini\footnote{Electronic address: 
Massimo.Giovannini@ipt.unil.ch }}

\address{{\it Institute of Theoretical Physics, University of Lausanne}}
\address{{\it BSP-1015 Dorigny, Lausanne, Switzerland}}

\maketitle

\begin{abstract}
Static solutions of the higher dimensional Einstein-Hilbert 
gravity  supplemented by quadratic curvature self-interactions
are discussed  in the presence of  hedgehog configurations
along the transverse dimensions.  
The quadratic part of the action is parametrized in terms of 
the (ghost-free) Euler-Gauss-Bonnet curvature invariant. 
Spherically symmetric profiles of the transverse metric 
admit exponentially decaying warp factors  both 
for positive and negative bulk cosmological constants. 
\end{abstract}
\vskip0.5pc
\centerline{Preprint Number: UNIL-IPT-00-20, September 2000 }
\vskip0.5pc
\noindent
\newpage

\renewcommand{\theequation}{1.\arabic{equation}}
\setcounter{equation}{0}
\section{Formulation of the problem} 

The idea that gravitational and gauge interactions can be unified in 
more than four space-time 
dimensions has been widely explored in various frameworks
\cite{kk}. 
More recently,  internal dimensions
have been analyzed in connection with possible alternatives 
of Kaluza-Klein compactification \cite{m1,m2} ( see also \cite{ak,viss} and 
\cite{dim}).

In this context \cite{m2,viss} our $(3+1)$-dimensional 
world might be interpreted as the internal space-time 
associated with a topological defect living in a higher dimensional manifold. 
This possibility has been scrutinized from different perspectives. 
New  solutions of the 
Einstein field equations (originally studied in the case of two 
transverse dimensions \cite{m2}) were obtained in the 
 case when
the number of transverse dimensions $n$ is equal to two  \cite{gs,ck,cp} 
or even larger than two \cite{vil,grs} (see also \cite{ran} for an earlier 
work on $n>2$ transverse dimensions).

The description of gravity  assumed in the 
discussion of \cite{vil,grs} relies on the Einstein-Hilbert action 
supplemented by a bulk cosmological constant and by matter sources 
(describing static topological defects living in the internal space). 
The aim of this paper is to discuss the 
scenario analyzed in \cite{vil,grs} in the framework of a different 
gravity theory where quadratic self-interactions are consistently 
included in the higher dimensional action.
The form of the gravity action considered in the present paper will then be
\footnote{A short remark concerning notations.
 $D= 4 + n$ is the dimensionality of the space-time ( $n$ labels the number 
of transverse dimensions). Riemann and Ricci  tensors
are defined as  $R^{\alpha}_{~\mu\beta\nu} = +\partial_{\beta} 
\Gamma^{\alpha}_{\mu\nu} +...$, $R_{\mu\nu}=R^{\alpha}_{~\mu\alpha\nu}$
($\Gamma^{\alpha}_{\mu\nu}$ is the Christoffel connection) and the signature 
of the metric is [-,+,+,+,...]).} 
\begin{equation}
S_{\rm g} = \int d^{D}x \,\sqrt{- g}\, \biggl[ -\frac{M^{D-2}}{2}\, R 
+ \Lambda_{D} 
- \alpha' R_{\rm EGB}^2\biggr],
\label{action}
\end{equation}
where $R$ is the curvature scalar in $D$-dimensions, $\Lambda_{D}$ is the 
bulk cosmological constant and $R_{\rm GB}^2$ is the quadratic part 
of the action which we choose to be in the form of the Euler-Gauss-Bonnet 
(EGB) combination:
\begin{equation}
R_{\rm EGB}^2 = R^{\beta\mu\alpha\nu}R_{\beta\mu\alpha\nu} - 4\, R^{\mu\nu}
R_{\mu\nu} + R^2.
\label{EGB}
\end{equation}
The coupling constant appearing in front of the EGB invariant 
has dimensions $\alpha' = [M^{D-4}]$. 
In the present analysis the full $D$-dimensional 
space-time will have a $(3+1)$-dimensional part and a transverse part 
formed by $n$  dimensions so that, dimensionally,
$\alpha'=[M]^{n}$.
Convincing  motivations coming from different 
theoretical ideas conspire towards 
the  EGB combination \cite{zw,des1} as a possible 
parameterization of quadratic self-interactions in higher 
dimensions.
In four dimensions the EGB combination is a topological term: 
its contribution to the 
action can be rearranged in a perfect four divergence which does not 
contribute to the classical equations of motion. In four dimensions 
the EGB combination coincides with the Euler invariant. 
In dimensions higher than 
four the EGB combination is not a total divergence and it does contribute 
to the classical equations of motion with terms containing (at most) 
a second derivative of the metric. 
Needless to say that, from a cosmological point of view, 
 the absence of derivatives higher 
than the second makes tractable (without the addition of extra initial 
conditions) 
the problem of the dynamical evolution \cite{mg,kal}. 
In a different perspective the EGB combination leads to ghost-free nontrivial 
gravitational self-interactions for $D>4$ \cite{zw,des1}. 
Quadratic terms in the gravitational action can naturally 
appear for diverse reasons. In string theory 
the (tree-level) low energy effective action is normally supplemented 
by two kinds of expansions: the expansion in the dilaton coupling (leading to 
higher genus correction) and the expansion in the string tension 
$\alpha'$ which 
involves the appearance of quadratic terms in the action \cite{gsw}
\footnote{ In our discussion 
we will assume that the dilaton coupling is frozen as assumed in \cite{RS}. 
See however 
\cite{mav} for a recent discussion of five-dimensional domain wall solution 
in the presence of gravitational self-interactions with static dilaton 
field.}.
In supergravity the EGB is required for the 
supersimmetrization  of the 
Lorentz-Chern-Simons term.  Higher dimensional theories with gravitational  
self-interactions were also investigated in the context of spontaneous 
compactification \cite{fr} induced by quadratic curvature 
corrections \cite{wett}.

Consider now a $4+n$-dimensional metric consistent with four-dimensional 
Poincar\'e invariance 
(of the type proposed in \cite{m2}) whose line element can be written as:
\begin{equation}
ds^2 =  g_{\mu\nu} dx^{\mu} dx^{\nu}= 
\sigma(y^a)\eta_{AB} dx^A dx^{B} + \gamma_{a b} dy^a d y^b,
\end{equation}
where $\sigma(y^a)$ is a conformal factor which only depends 
upon the the internal coordinates $y^a$; $\gamma_{a b}(y)$ parameterizes the 
metric 
of the internal space and $\eta_{A B}$ is the four-dimensional Minkowski metric
\footnote{Greek indices run over the whole $n +4$ space.
 Latin (capital) indices 
run in the four-dimensional world whereas small Latin letters label indices 
in the 
extra space.}.

In \cite{gs} solutions of the Einstein field 
equations have been obtained with a bulk (negative) cosmological constant 
and for local (string-like) defect along the 
(two-dimensional) transverse space. 
In \cite{vil,grs} higher dimensional transverse spaces have been 
studied with particular attention to global (hedgehog) 
defects  present in the transverse space. 

The aim of the present paper is twofold. On one hand 
possible solutions of the Einstein-Hilbert theory (supplemented 
by quadratic self-interactions) will be studied 
in the case of a higher dimensional transverse 
space. On the other hand useful generalizations of previous results will be 
provided. The present analysis is, therefore, complementary  to the ones 
of Ref. \cite{vil,grs}.
Gravitational self-interactions have been studied in the case of one or two
transverse dimension \cite{mav,q}, but not in the case of higher dimensional 
(global) defects like the ones discussed in the present paper.

The plan of the present paper is then the following. In Section II 
the basic equations will be derived. 
In Section III solutions with exponentially decreasing warp factors  
and spherically symmetric transverse metric will be presented.
Section IV deals with some possible extensions of the results with 
particular attention to the case of a bulk electromagnetic field.
Finally section V contains the concluding remarks. 

\renewcommand{\theequation}{2.\arabic{equation}}
\setcounter{equation}{0}
\section{Basic equations} 
The total action of the problem at hand 
is given by the sum of the gravity action 
of Eq. (\ref{action}) supplemented by the appropriate matter 
sources:
\begin{equation}
S= S_{\rm g} + S_{\rm m}.
\end{equation}
By functionally deriving the action with respect to the metric we obtain 
the corresponding equations of motion:
\begin{eqnarray}
&&R_{\mu\nu} -\frac{1}{2} g_{\mu\nu} R = - \frac{\Lambda}{M^{n + 2}} 
g_{\mu\nu} + 
\frac{T_{\mu\nu}}{M^{n + 2}} + \frac{2 \alpha'}{M^{n + 2}} {\cal Q}_{\mu\nu},
\nonumber\\
&& \nabla_{\mu} T^{\mu\nu} =0,
\label{eq}
\end{eqnarray}
where $T_{\mu\nu}$ contains the contribution of the matter sources  (obeying
covariant conservation laws) and  
where ${\cal Q}_{\mu\nu}$ contains the contribution of the EGB which 
turns out to be \cite{des1} 
\begin{equation}
{\cal Q}_{\mu\nu} = \frac{1}{2} g_{\mu\nu} R_{\rm EGB}^2 - 2 R~R_{\mu\nu} + 
4 R_{\mu\alpha}~ R^{\alpha}_{~~\nu} 
+ 4 R_{\alpha\beta}~ R^{\alpha\,\,\,\beta}_{\,\,\,\mu\,\,\,\nu}
- 2 R_{\mu\alpha\beta\gamma} ~R_{\nu}^{~~\alpha\beta\gamma}. 
\label{eq2}
\end{equation}
In Eq. (\ref{eq2}) it  has been assumed that the physical 
cosmological constant vanishes. In other words, the condition 
\begin{equation}
R^{(4)}_{A B} - \frac{1}{2}g_{A B} R^{(4)} = 0,
\end{equation}
will be implemented.
A hedgehog living along three transverse dimensions will be described through 
a triplet of scalars $(\phi^{1},\,\,\phi^{2},\,\,\phi^{3})$ whose potential 
can be chosen to be:
\begin{equation}
V(\phi) = \frac{\lambda}{4}( \phi^{a} \phi^a - v^2), 
\end{equation}
where, as previously mentioned, the (small) Latin indices run over the 
transverse directions. Outside the core the hedgehog ansatz
\begin{equation}
\phi^a(\rho) = v\,\,\frac{\rho^{a}}{\rho}
\end{equation}
will be considered ($ \rho^{a}/\rho$ is the unit vector along the 
transverse space). The three-dimensional 
metric of the transverse space will be chosen to be 
spherically symmetric so that the full (seven-dimensional) line element can 
be written as 
\begin{equation}
ds^2 = \sigma(\rho) \eta_{A B} dx^{A} d x^{B} + d\rho^2 + 
\gamma(\rho) [ d\vartheta^2 + \sin^2{\vartheta} d\varphi^2],
\label{m2}
\end{equation}
with $\eta_{AB}$ being the ordinary $(3+1)$-dimensional 
flat space metric $[-1,~+,~,+,~+]$ with signature ``mostly plus''. 
Notice that $\rho\geq 0$.  
Consequently, the components of the energy-momentum tensor will be:
\begin{eqnarray}
&&T_{\rho}^{\rho} = 
-\frac{v^2}{\gamma},\,\,\,\,\,\,T_{A}^{B} = - \frac{v^2}{\gamma}
\delta_{A}^{B},
\nonumber\\
&&T_{\vartheta}^{\vartheta} = T_{\varphi}^{\varphi} = 0.
\end{eqnarray} 
The purpose of the following Section will be to study some 
particular solution of the system of equations (\ref{eq}) with specific 
attention to the role played by the quadratic corrections of Eq. (\ref{eq2}).

\renewcommand{\theequation}{3.\arabic{equation}}
\setcounter{equation}{0}
\section{ Solutions with exponential warp factors} 
Defining ${\cal H} = [ \ln{\sigma}]'$, ${\cal F} = [\ln{\gamma}]'$, 
the EGB invariant can be written as 
\begin{eqnarray} 
R_{EGB}^2 &=& \biggl[ - 4 {\cal F}^3 {\cal H} - 
\frac{4}{\gamma} ({\cal H}^2+ 4 \frac{\sigma''}{\sigma}) 
+ {\cal F}^2 {\cal H}^2 -
\frac{9}{2} {\cal H}^4 + 8 {\cal F}{\cal H} \frac{\gamma''}{\gamma} 
\nonumber\\
&+& 
4 {\cal F}^2 \frac{\sigma''}{\sigma}   
+ 24 {\cal F} {\cal H} \frac{\sigma''}{\sigma} 
+ 12 {\cal H}^2 \biggl(\frac{\sigma''}{\sigma} + \frac{\gamma''}{\gamma}
\biggr)
\biggr].
\end{eqnarray}
Consequently, using the line element  (\ref{m2}), 
Eqs. (\ref{eq})--(\ref{eq2}) become, in components,
\begin{eqnarray}
&&{\cal F}' + \frac{3}{2} {\cal H}' + \frac{3}{2} {\cal H}{\cal F} 
+ \frac{3}{2} {\cal H}^2 
+ \frac{3}{4}{\cal F}^2 
- \frac{1}{\gamma} = - \biggl( \frac{\Lambda}{M_{7}^5} 
+ \frac{v^2}{\gamma~~M_{7}^5}\biggr) 
 +\frac{ 2 \alpha'}{M_{7}^5} {\cal G}_{0}(\rho),
\label{002}\\
&& \frac{{\cal F}^2}{4} + \frac{3}{2} {\cal H}^2 
+ 2 {\cal H}{\cal F} - \frac{1}{\gamma}
=  - \biggl( \frac{\Lambda}{M_{7}^5} 
+ \frac{v^2}{\gamma~~M_{7}^5}\biggr)  
+\frac{ 2 \alpha'}{M_{7}^5} {\cal G}_{\rho}(\rho) ,
\label{rr2}\\
&& 2 {\cal H}' + \frac{{\cal F}'}{2} 
+ \frac{5}{2}{\cal H}^2 + \frac{1}{4} {\cal F}^2  + {\cal H}{\cal F} 
=  -  \frac{\Lambda}{M_{7}^5}  
+ \frac{ 2 \alpha'}{M_{7}^5} {\cal G}_{\vartheta}(\rho),
\label{thth2}\\
&& 2 {\cal H}' + \frac{{\cal F}'}{2} + \frac{5}{2} {\cal H}^2 
+ \frac{1}{4} {\cal F}^2 + {\cal H}{\cal F} =
- \frac{\Lambda}{M_{7}^5}  
+ \frac{ 2 \alpha'}{M_{7}^5} {\cal G}_{\varphi}(\rho),
\label{phph2}
\end{eqnarray}
where
\begin{eqnarray}
{\cal G}_{0} (\rho)&=&\biggl[\frac{3}{2} {\cal H}{\cal F}^3 
+ \frac{9}{2} {\cal H}^3 {\cal F} 
+ \frac{3}{4} {\cal H}^4 + 3 {\cal H} {\cal F} {\cal F}' 
+ \frac{9}{2} {\cal H}^2 {\cal F}^2 
\nonumber\\
&+& \frac{3}{2} {\cal F}^2 {\cal H}' 
+ 6 {\cal H}{\cal F} {\cal H}' + \frac{3}{2} {\cal H}^2 {\cal H}' 
- \frac{6}{\gamma} ( {\cal H}^2  + {\cal H}') \biggr],
\label{g0}\\
{\cal G}_{\rho}(\rho) &=&  
 \biggl( \frac{3}{4} {\cal H}^4 + 6 {\cal F}{\cal H}^3 + \frac{9}{2} 
{\cal F}^2 {\cal H}^2 - \frac{6}{\gamma} {\cal H}^2\biggr),
\label{gr}\\
{\cal G}_{\vartheta}(\rho)&\equiv& {\cal G}_{\varphi} (\rho) 
=  \biggl( \frac{ 3}{2}  {\cal H}^2 {\cal F}^2 + 3 {\cal H}^2 {\cal F}' + 
6 {\cal H}{\cal F} {\cal H}' + 6 {\cal H}^3 {\cal F} 
+ \frac{15}{4} {\cal H}^4 + 6 {\cal H}^2{\cal H}'\biggr).
\label{gth}
\end{eqnarray}
Eqs. (\ref{002})--(\ref{phph2}) are, respectively,
 the $(00)$, $(\rho~\rho)$, $(\vartheta~\vartheta)$ and 
$(\varphi~\varphi)$ components of Eq. (\ref{eq}). 
Notice that 
the ($\vartheta$,$\vartheta$) and ($\varphi$,$\varphi$) components 
of Eq. (\ref{eq}) lead to the same equation.

If $\alpha'=0$ (i.e. in the absence of quadratic corrections) 
a particular solution of the previous system is \cite{grs}
\begin{equation}
\sigma{\rho} = e^{ - \frac{\rho}{L}}, \,\,\, \gamma(\rho) = \gamma_{L} 
= {\rm constant}. 
\end{equation}
Inserting ${\cal H} 
= - 1/L$ and ${\cal F} =0$, into Eqs. (\ref{002})--(\ref{phph2}) the 
relations
\begin{eqnarray}
&&L= \sqrt{ - \frac{ 5 M_{7}^5}{2 \Lambda} },
\label{L0}\\
&&\gamma_{L} = L^2 \biggl( \frac{v^2}{M_{7}^5} -1\biggr),
\label{gamma0}
\end{eqnarray}
can be obtained. In eq. (\ref{L0})
 $ \Lambda <0$ and $v^2/M_{7}^{5}$ is a dimensionless quantity.
By now tuning $v^2 = M_{7}^5$ a different solution can be obtained, namely 
\begin{equation}
{\cal F} = {\cal H}= -\frac{1}{L} ,\,\,\, \sigma(\rho) = q~\gamma(\rho),
\label{ans}
\end{equation}
with
\begin{equation}
L= \sqrt{ - \frac{15}{4}\frac{M_{7}^5}{\Lambda}},
\end{equation}
also in this case the cosmological constant needs to be negative. 

If $\alpha'\neq 0$ new solutions can be obtained. 
In particular, assuming that $ {\cal H} = -1/L$ and 
that $ {\gamma}_{L}= {\rm constant}$,
consistency with Eqs. (\ref{002})--(\ref{phph2}) requires
\begin{eqnarray}
&&\frac{3}{2~~L^2 } - \frac{1}{\gamma_{L}} = - \frac{\Lambda}{M^5_{7}} 
- \frac{ v^2}{ \gamma_L~~M_{7}^5}
+ \frac{ 2 \alpha'}{M_{7}^5}\,\, \biggl( \frac{ 3}{ 4 ~~L^4} 
- \frac{6}{ L^2\, \gamma_{L}}\biggr),
\nonumber\\
&& 2\, \frac{ \Lambda}{M^5_{7}}\, L^4 \,\,+\,\, 5\, L^2\,\, -\,\, 15\,
 \frac{\alpha'}{M_{7}^5} =0,
\end{eqnarray}
whose solution is 
\begin{eqnarray}
&&\gamma_{L}(L) = \frac{ L^4 (  M_{7}^5 - v^2) -
 12 \,\alpha' \,L^2}{L^4 \Lambda 
+ \frac{3}{2} (L^2 M_{7}^5 - \alpha')},
\label{gammaa}\\
&& L^2 =  -\frac{5}{4} \frac{M_{7}^5}{\Lambda} \biggl[1 \mp 
\sqrt{ 1
 + \frac{24}{5} \frac{\Lambda~\alpha'}{M_{7}^{10}}}\biggr].
\label{Lmp}
\end{eqnarray}
In the limit $\alpha' \rightarrow 0$, the expression 
for  $\gamma_{L}$ obtained in Eq. (\ref{gamma0}).
is formally recovered.
 Define, for sake of simplicity,
 $\Lambda = \Lambda_{+}$ (if $\Lambda > 0$) and 
$ \Lambda = - \Lambda_{-}$ (if $\Lambda < 0$). 
Then,  from Eq. (\ref{Lmp}), 
\begin{eqnarray}
&&L_{+} = \biggl\{ -\frac{5}{4} \frac{M_{7}^5}{\Lambda_{+}} \biggl[1 -
\sqrt{ 1 
+ \frac{24}{5} \frac{\Lambda_{+}~\alpha'}{M_{7}^{10}}}\biggr]\biggr\}^{1/2},
\label{plus}\\
&& L_{-} =  \biggl\{ \frac{5}{4} \frac{M_{7}^5}{\Lambda_{-}} \biggl[1 +
\sqrt{ 1 
- \frac{24}{5} \frac{\Lambda_{-}~\alpha'}{M_{7}^{10}}}\biggr]\biggr\}^{1/2}.
\label{minus}
\end{eqnarray}

From Eq. (\ref{Lmp}), an exponentially decreasing warp factor can be 
obtained either 
from $L_{+}$ ( with $\Lambda > 0$) or from $L_{-}$ 
(with $\Lambda <0$).
For positive bulk cosmological constant $L_{+}$ is always defined. 
In the case of negative cosmological constant $L_{-}$ 
is defined only if 
\begin{equation}
\frac{\alpha' \Lambda_{-}}{M_{7}^{10}} < \frac{5}{24}.
\label{bound}
\end{equation}
Thus, exponentially   decreasing warp factors 
are possible for any positive cosmological 
constant and for negative cosmological constants whose absolute value
satisfies Eq. (\ref{bound}).

In the limit $\alpha'\rightarrow 0$ we get  
$L_{-} \rightarrow 0$ and $\gamma_{L}(L_{-}) \rightarrow 0$. 
On the contrary, for $\alpha'\rightarrow 0$,  
$L_{+} \rightarrow \sqrt{ -5 M_{7}^5/2 \Lambda}$ and the 
solution in the absence of EGB invariant is reproduced since 
$\gamma_{L}(L_{+})$ exactly equals 
the relation derived in Eq. (\ref{gamma0}). 
Therefore, the inclusion of quadratic 
self-interactions has a twofold effect. On one hand it generalizes 
the tree-level solution.
On the other hand it introduces a new solution whose limit 
(for $\alpha' \rightarrow 0$ ) is 
the (seven-dimensional) Minkowskian space.  

Given a specific $L_{\pm}$, $\gamma_{L}(L_{\pm})$ can be determined 
through Eq. (\ref{gammaa}).
Since the signature of the metric should be preserved
 $\gamma_{L}(L_{\pm})$ is required to be positive definite. 
Thus Eq. (\ref{gammaa}) 
together with Eqs. (\ref{plus})--(\ref{minus}) imposes a physical 
bound for the scalar vacuum expectation value $v^2$. 
Using the equation satisfied by $L_{\pm}$ in the denominator 
of the relation (\ref{gammaa}) we 
can see that $\gamma_{L}(L_{\pm}) > 0 $ if
\begin{equation}
v^2 > \frac{L_{\pm}^2 M_{7}^5 - 12 \alpha'}{L_{\pm}^2}.
\end{equation}

A different solution to Eqs. (\ref{002})--(\ref{phph2}) can be obtained 
by tuning $v^2 = (L^2 M_{7}^5 - 12 \alpha')/L^2$.  In this case the 
$1/\gamma$ factors disappear from the equations of motion whose 
integration can be performed  by assuming that 
$\sigma \propto \gamma$ or, in terms of their logarithmic 
derivatives, that ${\cal H}= {\cal F}$. 
In this case the equations of motion 
(\ref{002})--(\ref{phph2}) can be consistently 
solved assuming an exponentially decreasing warp factor. 
Following the same steps outlined in the previous case $L$
 can be found, and it turns out to be 
\begin{equation}
L^2 = - \frac{15 M^5}{8 \Lambda}( 1 \mp \sqrt{ 1 
+ \frac{32 \alpha' \Lambda}{5 ~M_{7}^{10}}}).
\label{eqL}
\end{equation}
As in the previous the limit $\alpha' \rightarrow 0$ can be discussed. 
One limit gives back the tree-level solution the other limit 
gives the (seven-dimensional ) Minkowski space. 
Notice that, again,  when $\alpha'\neq 0$ the exponential 
decrease in the warp factor can 
be achieved both with positive and negative bulk cosmological constant. 

This last solution has a curvature singularity for $\rho \rightarrow \infty$. 
This aspect can be illustrated by computing the 
EGB invariant on the obtained solutions:
\begin{equation}
R_{EGB}^2 = \frac{105}{2\,L^4} - \frac{20\,e^{\frac{\rho}{L}}}{q\,L^2}
\end{equation}
where $q$ is the proportionality factor between $\sigma $ and $\gamma$ 
as it appears in Eq. (\ref{ans}).
In spite of the singular character of the geometry the action is finite. 
This solution has already a singularity of the same type already without 
curvature self-interactions and the addition of the EGB contribution 
to the action does not modify this aspect.

\renewcommand{\theequation}{4.\arabic{equation}}
\setcounter{equation}{0}
\section{Possible Extensions: electromagnetic fields} 

In the present section the self-interacting gravitational 
field will be studied in the presence of a bulk electromagnetic field.
The matter action will take then the form 
\begin{equation}
S_{\rm m} = - \frac{1}{4} \int d^7 x \sqrt{-g} F_{\mu\nu} F^{\mu\nu}. 
\end{equation}
The equations of motion and the Bianchi identities can be written as 
\begin{eqnarray}
&&\partial_{\mu}\biggl( \sqrt{-g} F^{\mu\nu}\biggr) =0,
\nonumber\\
&& \partial_{\mu}\biggl( \sqrt{-g} \tilde{F}^{\mu\nu}\biggr) =0.
\end{eqnarray}
A specific magnetic field configuration will then be chosen, namely 
the electromagnetic field will be completely polarized along 
the transverse dimensions and its only non-vanishing component will be 
\begin{equation}
F_{\vartheta\varphi} = Q \sin{\vartheta}.
\end{equation}
where $Q$ is a constant and, as usual,
 ($\vartheta$,$\varphi$) label the radial 
coordinates of the transverse metric.
Taking now into account that the relevant components of the energy-momentum 
tensor are 
\begin{equation}
T_{A}^{B} = -\frac{1}{2} \frac{Q^2}{\gamma^2} \delta_{A}^{B},\,\,\,\,
T_{\rho}^{\rho} =  -\frac{1}{2} \frac{Q^2}{\gamma^2},\,\,\,\,
T_{\vartheta}^{\vartheta} = T_{\varphi}^{\varphi}= 
\frac{1}{2} \frac{Q^2}{\gamma^2},
\end{equation}
we can write the equations of motion
\begin{eqnarray}
&&{\cal F}' + \frac{3}{2} {\cal H}' + \frac{3}{2} {\cal H}{\cal F} 
+ \frac{3}{2} {\cal H}^2 
+ \frac{3}{4}{\cal F}^2 
- \frac{1}{\gamma} = - \biggl( \frac{\Lambda}{M_{7}^5} 
+ \frac{Q^2}{2~\gamma^2~~M_{7}^5}\biggr) 
 +\frac{ 2 \alpha'}{M_{7}^5} {\cal G}_{0}(\rho),
\label{003}\\
&& \frac{{\cal F}^2}{4} + \frac{3}{2} {\cal H}^2 
+ 2 {\cal H}{\cal F} - \frac{1}{\gamma}
=  - \biggl( \frac{\Lambda}{M_{7}^5} 
+ \frac{Q^2}{2~\gamma^2~~M_{7}^5}\biggr)  
+\frac{ 2 \alpha'}{M_{7}^5} {\cal G}_{\rho}(\rho) ,
\label{rr3}\\
&& 2 {\cal H}' + \frac{{\cal F}'}{2} 
+ \frac{5}{2}{\cal H}^2 + \frac{1}{4} {\cal F}^2  + {\cal H}{\cal F} 
=  -  \frac{\Lambda}{M_{7}^5}  + \frac{Q^2}{2~\gamma^2~~M_{7}^5}+
 \frac{ 2 \alpha'}{M_{7}^5} {\cal G}_{\vartheta}(\rho),
\label{thth3}
\end{eqnarray}
[the ${\cal G}$ functions being the same as the ones 
of Eqs. (\ref{g0})--(\ref{gth})].

A solution of the form  
\begin{equation}
{\cal H} = -\frac{1}{L},\,\,\, \gamma= \gamma_{L},
\end{equation}
will then be analyzed ($L$ and $\gamma_{L}$ are, as in the previous Section,
 two positive  constants). In the absence of curvature 
corrections this solution has been discussed in \cite{grs}.

From Eqs. (\ref{003})--(\ref{thth3}) the following relations can then be 
obtained
\begin{eqnarray}
&& 9 ~ d ~ x^4 + 2 y ( 1 - 6~d~ x^2) - 8 ~x^2 - 4~a=0,
\label{alge}\\
&& 3 ~d~ x^4 - y~ ( 1 - 6~ d~x^2) - x^2 + b~ y^2 =0,
\label{alge1}
\end{eqnarray}
where, for simplicity, we define
\begin{equation}
x= 1/L,\,\,\, y = 1/\gamma_{L},\,\,\,\ a= \frac{\Lambda}{M_{7}^5},\,\,\,
b=\frac{Q^2}{M_{7}^5},\,\,\,\, d = \frac{2 \alpha'}{M_{7}^5}.
\end{equation}
Eqs.(\ref{alge})--(\ref{alge1}) 
 are difficult to solve, in general. They can be 
numerically solved for specific values of the parameters. 
Also in this case both positive and negative cosmological constants 
lead to exponentially decreasing warp factors. 

We can illustrate this point in the following way. Working in (natural) 
gravitational units (i.e. $M_{7} = 1$)  
 solutions of the two algebraic equations (\ref{alge}) and (\ref{alge1}) 
can be analyzed for the cases where the roots are simultaneously positive.
Different specific values 
of $b$ and $d$ can be chosen. Keeping $d$ and $b$ fixed, the value of 
$a$ can be 
changed from positive to negative. Both for positive and negative 
$a$ (corresponding to positive and negative cosmological constants), 
pairs of positive roots of Eqs. (\ref{alge})--(\ref{alge1})  can be found. 
For instance, choosing $b =0.1$ and $d =0.1$, one gets two positive 
pairs of roots both for positive and negative $a$
\begin{equation}
(y,x)\equiv  (5.006,~~1.036),\,\,\,\, {\rm for}\,\, a= -1,\,\,
(y,x)\equiv  (6.961,~~0.792),\,\,\,\, {\rm for}\,\, a= 1,\,\,
\end{equation}
Different sets of parameters can be chosen with similar conclusions.
\begin{figure}
      \centerline{\epsfxsize = 9 cm  \epsffile{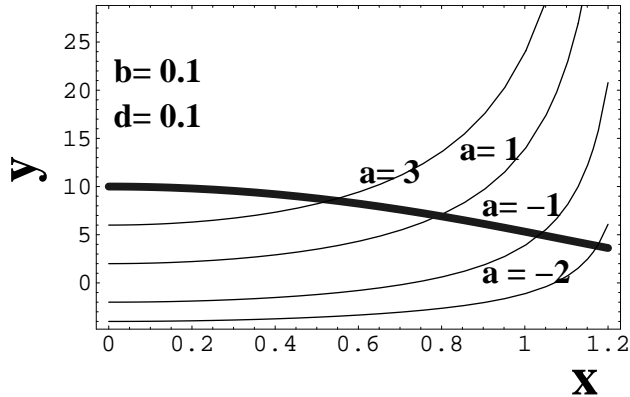}} 
\caption[a]{The system of Eqs. (\ref{alge})--(\ref{alge1}) is illustrated 
for $b= d = 0.1$ (in units $M_{7}=1$ ) for different values of $a$.}
\label{f1}
\end{figure}
The value of $a$ can be changed (for fixed $d$ and $b$) and, in this way, the 
space of the possible solutions can be explored. 
In Fig. \ref{f1} this point is illustrated. With the thin lines 
the relation between $y$ and $x$ is plotted (for different values of $a$)
 according to Eq. (\ref{alge}). 
With the thick line the relation 
between $y$ and $x$ following from Eq. (\ref{alge1}) is reported. 
In Fig. \ref{f1} $b=0.1$ and $d=0.1$ are fixed. 
The intersections of the thin lines with the thick 
line will give the (real) roots of the system 
for different values of $a$. We can observe that the intersections, in 
the example of Fig. \ref{f1}, all lead to positive roots.
Notice that 
Eq. (\ref{alge1}) does not contain $a$ (i.e. the cosmological constant) 
 but only $b$ and $c$ (i.e.  $Q$ and $\alpha'$). 
This argument shows that positive and negative cosmological 
constants lead to consistent solutions. In order to fully explore 
the space of the solution the values of $b$ and $d$ should change.   
\renewcommand{\theequation}{5.\arabic{equation}}
\setcounter{equation}{0}
\section{Concluding remarks} 
In this paper solutions of the Einstein field equations with 
EGB self-interactions have been studied when a hedgehog configuration 
is present in the (transverse) higher dimensional space. 
The  occurrence of exponentially decreasing warp factors has 
been pointed out. If quadratic self-interactions are absent, 
exponentially decreasing warp factors arise only for 
negative bulk cosmological constant. 
If quadratic self-interactions are present  such a behavior
is allowed both for positive and negative cosmological constants. 

Two specific classes of solution have been examined. In the first class 
of solutions the  functions describing the variation of the metric 
along the transverse coordinates are constant.
 In the second class of solutions 
the metric functions of the transverse space change with the distance 
from the core of the hedgehog. In this case the space-time is 
singular at large distances even in the presence of the EGB self-interactions. 

We also discussed similar solutions in the case of a specific 
magnetic field configuration completely polarized along 
the transverse dimensions. Also in this case exponentially 
decreasing warp factors are stable towards quadratic self-interactions and can
arise for positive and negative cosmological constants.  

\section*{Ackwoledgments} 

The author would like to thank M. E. Shaposhnikov for interesting
discussions.

\end{document}